\documentclass{elsart}
\journal{Computer  Physics Communications}
\usepackage{graphicx,subfigure,amssymb,amsmath}

\begin{document}
\begin{frontmatter}
\title{Numerical differentiation of experimental data: local versus global methods}
\author{Karsten Ahnert}\ead{kahnert@stat.physik.uni-potsdam.de},
\author{Markus Abel}\ead{markus@stat.physik.uni-potsdam.de}
\address{Institute of Physics, University of Potsdam, 14415 Potsdam,
  Germany}

\begin{abstract}
  In the context of the analysis of measured data, one is often faced with the
  task to differentiate data numerically. Typically, this occurs when measured
  data are concerned or data are evaluated numerically during the evolution of
  partial or ordinary differential equations. Usually, one does not take care
  for accuracy of the resulting estimates of derivatives because modern
  computers are assumed to be accurate to many digits. But measurements yield
  intrinsic errors, which are often much less accurate than the limit of the
  machine used, and there exists the effect of ``loss of
  significance'', well known in numerical mathematics and computational
  physics. The problem occurs primarily in numerical
  subtraction, and clearly, the estimation of derivatives involves 
  the approximation of differences.  In this article, we discuss several
  techniques for the estimation of derivatives. As a novel aspect, we divide
  into local and global methods, and explain the respective shortcomings. We
  have developed a general scheme for global methods, and illustrate our ideas
  by spline smoothing and spectral smoothing. The results from these less
  known techniques are confronted with the ones from local methods. As typical
  for the latter, we chose Savitzky-Golay filtering and finite differences.
  Two basic quantities are used for characterization of results: The variance
  of the difference of the true derivative and its estimate, and as important
  new characteristic, the smoothness of the estimate. We apply the different
  techniques to numerically produced data and demonstrate the application to
  data from an aeroacoustic experiment.  As a result, we find that global
  methods are generally preferable if a smooth process is considered.
  For rough estimates local methods work acceptably well.
\end{abstract}

\end{frontmatter}

\section{Introduction}
The estimation of derivatives from numerical data is a classical problem which
occurs in many problems of data analysis \cite{Kantz-Schreiber-97}.
Applications range from biology
\cite{Damico-Ferrigno-92,Schaefer-Rosenblum-99}, and chemistry
\cite{Kiss-Hudson} to a variety of problems in physical applications
\cite{Abel-Ahnert-Bergweiler-07,PIV-01,Conway-private}; various mathematical
aspects are discussed in
\cite{Cullum-71,Anderssen-Bloomfield-74,Anderssen-Hegland-99,Hanke-Scherzer-2001}.
Surprisingly, experimenters and numerical data analysts often use very crude
techniques for the estimation of derivatives and the topic is not discussed in
depth in the physics community.  Despite the relatively simple strategies to
enhance results for numerical differentiation, improved techniques are rarely
used.  With this contribution, we formulate a general scheme to distinguish
the existing methods and discuss in detail their differences. Our emphasis is
on {\it accuracy} and {\it smoothness}, so wherever one is not bothered with
very accurate estimation or with very noisy data, our text is understood to be
informative.  However, in critical applications with a high standard on
accuracy of the data, this paper can guide along the numerical subtleties of
the estimation of derivatives with a minimum loss of accuracy. The task we try
to analyze is to differentiate numerically a data series with given accuracy
in such a way that the loss of accuracy is minimal and the smoothness of the
result, defined later in this text, is maximal.  In the following, we present
typical techniques and algorithms in a computation-oriented way, referring to
the literature for mathematical details.

Whereas mathematically, the derivative of a function is obtained by a limit
process involving infinitesimal calculus, this can never be realized for data
measured by digital equipment due to the intrinsic discretization of the data. In
addition, real measurements yield data with noise, either due to the
device properties or due to finite resolution, e.g., when sampling the data.
An estimate for a derivative of some data $y(x)$ with respect to the argument
$x$ has to cope with these two restrictions. Noise and discretization effects
yield errors in the data $y$, the finite sampling interval does not allow the
limit $\Delta x \to 0$ and thus produces numerical errors, too.

Typically, to obtain an estimate for the derivative $y'(x)$, one tries to
approximate the measured data $y$ best (in an exactly specified sense). One
then hopes that the derivative found from this approximation is as well best.
In this article, we want to obtain maximal accuracy, but as well the smoothest
result possible. One can divide the existing techniques into local and global
strategies. For the first, one fits or interpolates a function through the
data points locally, as for a running window; in the second case, the fit or
interpolation is global. A typical example for global fits are {\it smoothing
  splines} as an enhancement of the well-known interpolating splines
\cite{Wahba-90}. In contrast to interpolation a spline is put through {\it
  many} data fulfilling a least-squares condition. This way, one can take into
account a natural property of physical variables - their smoothness, which is
very often assumed for theoretical or experimental reasons. Taken into account
this criterion, differences between methods can be clearly noticed. The
quantification of the smoothness property is novel to the numerical
differentiation, we believe that it is an important and useful criterion.
Another quantification of the quality of the estimate than smoothness is the
least-squares error. A consistent formulation including a smoothness parameter
yields a minimization problem, mathematically known as Tikhonov regularization
\cite{Hanke-Scherzer-2001,Groetsch-84}.

Sometimes, a kernel smoother is applied before fitting the data. This is a
quite crude technique and we can not recommend this method, since two
regression methods are combined and one carries the disadvantages of both
methods. Furthermore, most methods work better with the original data \cite{NR}. But we
must also note, that some combinations of kernel smoothers and regression
methods might be successfully applied to some special problems.

In this article, we compare four different techniques: finite differencing
\cite{Trefethen-96}, the Savitzky-Golay filter \cite{NR},
smoothing splines \cite{Wahba-90} and a spectral technique
\cite{Abel-Ahnert-Bergweiler-07,NR}. Whereas the first two are local methods,
the latter are global ones. Local methods typically produce non smooth
functions, whereas for global methods the smoothness is controlled in a
well-defined way. We compare the methods by two numerically produced data sets
where the noise in the data can be controlled and the derivative is known
either analytically or numerically with high accuracy. Finally, we apply the
methods to a set of data from an aeroacoustic measurement
\cite{Abel-Bergweiler-Multhaupt-06}.

This article is organized as follows: In Section \ref{sec:estimate}, we give a
brief overview of the topic and details on local and global methods. Results
for the implementation of the methods are given in Section \ref{sec:results}
for the three above mentioned examples. The article concludes with Section
\ref{sec:conclusion}.

\section{Estimation of Derivatives}
\label{sec:estimate}
The estimation of derivatives is frequently encountered in the numerical
integration of ordinary differential equations \cite[Chp.~5,16]{NR}. A
function $f(x)$ is given and one has to compute its derivative.  For this
task, the typical estimators like finite difference schemes, spectral or
kernel methods are used
\cite{Trefethen-96,NR,Torcini-Frauenkron-Grassberger-97} in dependence on the
required accuracy and speed of the numerics. The key problem
concerning accuracy is the cancellation of digits with numerical subtraction,
mentioned above. An introductory analysis is given in \cite[Chp.~5]{NR}, a
more detailed exposure is found in \cite{NR,Goldberg-91,Stoer-Burlisch-93}.

A conceptually different problem arises if one wants to estimate derivatives
from measured data {\em without knowing the underlying function}. Then, one
needs first a ``good'' estimate for $f(x)$ and then another ``good'' method to
extract the derivative. What ``good'' means can vary heavily from data set to
data set. The algorithms and methods one uses to solve this problem depend on
the requirements the analyst imposes on the data. E.g., it might be
unimportant for an experimenter if his result is accurate up to $2$ or $10$
digits because he only needs a graphical representation of the results. On the
other hand, some applications, like differential embedding
\cite{Sauer-Yorke-Casdagli-91}, might require the knowledge of several
derivatives at high accuracy for further analysis of a dynamical system. Then,
usual techniques, as given in \cite{NR} very quickly meet their limits
regarding accuracy and  smoothness of a numerically calculated derivative.

In the mathematical literature, the problem has been analyzed as inverse
problem under numerical aspects \cite{Hanke-Scherzer-2001,NR,Wei-Hon-Wang-05}.
Here, we follow the rigorous ideas of \cite{Hanke-Scherzer-2001} and present a
classification of the schemes into global and local ones; we illustrate the
differences by typical examples.  Our focus is on the {\em smoothness} of the
estimated derivatives, because a description of a physical system typically
requires functions and derivatives to be smooth up to a certain order. For
example, most Hamiltonian systems are smooth systems, since their solutions
are mostly required to be $C^2$.

In the following, we consider a data series $y(x)$, measured at $N$ points
$(y_i,x_i)$ ($i=1,...,N$). Due to measurement accuracy and noise sources the
data have errors and we assume that our measured values consist of a
deterministic part $f(x)$ and a noisy part $\eta(x)$, such that the measured
data are $y(x)=f(x)+\eta(x)$. This already indicates that the true function
$f(x)$ should be found using regression techniques. Since $f$ is generically
nonlinear, linear regression \cite{NR} is not applicable and nonparametric
methods should be used; for an introduction see \cite{Haerdle-90}. Please note
that this is a much more complex problem than the estimation of parameters
$\alpha_i$, $i=1,\dots,n$ by iterative methods if the form of
$f(x,\alpha_1,\dots,\alpha_n)$ is known, as, e.g., the standard
Levenberg-Marquardt one \cite{NR,Levenberg-44,Marquardt-63}.  We assume further that $f$ is a
smooth function and that the data series $y_i(x_i)$ is given on a uniform
grid $x_i=a + i \Delta x$, $i=0,1,2,\dots$ of length $L=b-a$ in the interval
$a\le x_i \le b$. Now, we are interested in the estimate for the derivatives
$f'(x)=\mbox{d}f/\mbox{d}x$ from the measured data. Throughout this article we
will denote the estimate of the derivative by $\tilde{f}'(x)$ and accordingly the
estimate of the function by $\tilde{f}(x)$.

After having experimented with many data sets and methods, we found out that
the techniques commonly used can be distinguished in a simple way as local or
global approximation methods. A local method approximates the function $f(x)$
in some neighborhood $\mathcal U(x)$ of $x$, and does not yield information
beyond.  This means especially, that there can be jumps in the approximated
function between different $x$ values. A global method includes
information about all points in the estimate for $f(x)$ and thus jumps can be
avoided if smooth functions are used for the approximation.

The basic idea in all the methods is to approximate the function $f(x)$ with
the hope that then the derivative can be estimated.  More explicitly one
assumes the following: if the approximation $\tilde{f}$ is close to the
original function $f$, then the derivative $\tilde{f}'$ is also close to $f'$.
In mathematical terms, if $||f-\tilde{f}||=\mathrm{min}$, then
$||f'-\tilde{f'}||\approx\mathrm{min}$, with $||\cdot||$ a suitable norm.  The
estimation is best in the sense of the applied norm. Usually, the $L_2$ norm
is chosen and one has to solve a least-squares problem
\cite{NR,Levenberg-44,Marquardt-63}. Common methods yield functions
$\tilde{f}$ obtained either by local or global interpolation, or a fit
\cite{NR}. In an interpolation problem the resulting function is required to
go through the data points, i.e. $\tilde{f}(x_i)=y_i$; a fit produces the
function such that $\sum_{j} \|\tilde{f}(x_j)-y_j \|=min$.  Implicitly, that
means that $\tilde{f}(x_i)$ is not necessarily identical to $ y_i$. A local
method acts on a subset of the data ($j\in(i-n_l,i+n_r)$), with  $n_r, n_l \in
I\!\! N$ a global one on all
available data.  Most of the methods can be extended to give approximate
function values or derivatives not only at the points $x_i$ but on the whole
interval $a\le x\le b$. At the boundaries some methods are problematic
because, e.g., the statistic changes, or the problem is no longer well posed.

\subsection{Local Methods}
\label{sec:estimate:local}

Local methods work by fitting or interpolating for each $x_i$ a function
$\tilde{f}(x)$ on some sub--interval $I \ni x$ of the domain. Obviously, this
does not guarantee that the function is smooth, because the measurement errors
can yield jumps (from interval to interval). Probably the best known method is
given by {\it finite differences}; it results in a natural way when
differential operators are discretized. This technique has turned to a huge
field in connection with numerical integration of partial differential
equations \cite{Trefethen-96}.  Depending on the problem, different schemes
can be used and the choice of the right difference scheme can have enormous
impact on the result of an integration \cite{Ferziger-Peric-02}. The basic
idea is simple: the function $f$ is interpolated by a polynomial of order
$m$. For symmetric differencing, it is put through the points
$x_{i-k},\dots,x_i,\dots,x_{i+k}$, where $m=2k$ is a positive integer,
asymmetric schemes, like the well-known upwind schemes work in a similar
way. The derivative at the point $x$ is then the derivative of the
interpolated polynomial. Finite differencing is local, because the
approximation of the derivative depends only on the $2k+1$ points in the
neighborhood of $x_i$.  The polynomial can be obtained from Taylor expansion,
Pad\'e approximation or similar schemes \cite{Ferziger-Peric-02}.
The finite difference estimator for a Taylor expansion has the form
\cite{Trefethen-96}
\begin{equation}
\Big(\frac{\partial \tilde{f}}{\partial x}\Big)_i = \sum_{j=1}^k
\alpha_j \frac{y_{i+j}-y_{i-j}}{2j\Delta x}\textrm{,}
\label{eq:taylor_general}
\end{equation}
with coefficients
\begin{equation}
\alpha_j=2\cdot (-1)^{j+1} \left( k \atop k-j \right)
  \left/ \left( k+j \atop k \right) \right.\mathrm{.}
\label{eq:taylor_coef}
\end{equation}
For other schemes, different coefficients are used. In that case
(\ref{eq:taylor_coef}) or the right hand side of
(\ref{eq:taylor_general}) can contain derivatives. It is also possible
to vary the step width $\Delta x$. To do so, one  replaces $\Delta x$ by
$\delta x=l \Delta x$ ($l\in I\!\!N$) and $j$ by $j^\prime=lj$ in (\ref{eq:taylor_general}).

If the data are noise-free, the error $e=\| f^\prime-\tilde{f^\prime}\|$ in
(\ref{eq:taylor_general}) is of order $O(\Delta x^m)$. This means, with a fine
sampling one can arrive at accurate approximations. Numerically, however, one
is faced with the unavoidable problem of accuracy loss in numeric subtraction
due to the loss of significance of digits
\cite{Goldberg-91,Stoer-Burlisch-93}. Subtraction of two floating point
numbers is ill conditioned if the value of two numbers is approximately equal.
This is typical for the scheme (\ref{eq:taylor_general}).  Especially if
measurement noise is present, it can easily travel to the leading digits and
render the results meaningless; i.e.  finite differencing is a bad choice for
numerical differentiation. One can consider the problem as a trade-off between
numerical inaccuracy due to subtraction and analytical need for small values
of $\Delta x$ for the approximation (\ref{eq:taylor_general}) to hold
sufficiently well.

A consequent error analysis including the data accuracy $\delta=\sqrt{VAR
  (\eta) }$, with $\eta$ the measurement noise process, yields the error of a
first-order symmetric finite difference scheme ($k=1$) in
  Eq.~(\ref{eq:taylor_general}) \cite{Hanke-Scherzer-2001}:
\begin{equation}
e \sim  \delta x+\delta/\delta x\;.
\label{eq:error_fd}
\end{equation}
A minimal error $e\simeq 2 \sqrt{\delta}$ is found for $\delta x \sim
\sqrt{\delta}$. With given data of sampling interval $\Delta x \ll
\sqrt{\delta}$ one needs to discard some points to achieve the minimum and
uses $\delta x=l \Delta x$ This is not satisfying, because the information
contained in the left-out points in the interval $y_{i-l},..,y_{i+l}$ is
thrown away. One would like to use a scheme which has minimal error, but uses
all points with the intention to go beyond the bound $2\delta$ for the error
of the estimate.  

This can be achieved by using a fit of a polynomial of order $m<2k$ through
all data in the interval $(x_{i-k},x_{i+k})$. This method is known as {\it
  Savitzky-Golay-filtering} and is widely used in data analysis. The domain
has not to be symmetric (as in generalized finite differencing), one defines a
neighborhood by an interval $(i-n_l,i+n_r)$ with $n_l+n_r+1$ points, and
$n_l$ not necessarily equal to $n_r$. A linear regression is used to find the
best polynomial fit of order $m$ to the data, and the derivative is obtained
from the coefficients of the polynomial. This procedure is repeated for every
data point, like for a moving window. The key idea of Savitzky-Golay filtering
is the conservation of higher statistical moments. A simple
moving average always reduces the height of a local extremum. Due to the
mentioned conservation property,  the Savitzky-Golay filter shows this
reduction to a much less extent. Smoothness, however, is not guaranteed
and the derivative can be discontinuous, which is not desirable for an
estimate useful in physical problems where $f$ is typically required to be
smooth. A very nice and detailed discussion is found in the numerical recipes
\cite{NR}.

\subsection{Global Methods}
\label{sec:estimate:global}
Global methods yield an estimation $\tilde{f}(x)$, defined on the whole
interval $a\le x\le b$. Since the function is not known beforehand, it makes
sense to use a representation by some basis functions, $\phi_i(x)$, $i\in
I\!\!N$: $f(x)=\sum_j a_j \phi_j(x)$. The estimation shall be best in the
least squares sense, but as well smooth. Consequently a minimization problem
with a side condition for smoothness is formulated. The coefficients $a_j$ are
determined accordingly. The choice of the basis depends on the properties on
$f$, either known by prior knowledge oder imposed {\it a posteriori}. For instance,
it might be clear from the experiment that $f \in C^2$ is smooth (in that the
second derivative exists), or continuous only, or periodic on the interval, or
has other restrictions which are known to the modeler.
 
To quantify the global smoothness of a function $g(x)$ one uses its curvature
\cite{Hastie-Tibshirani-90}
\begin{equation}
s(g)=\int_a^b \| g''(x)\|^2 \d x \textrm{.}
\label{eq:sm_s}
\end{equation}
Maximizing the smoothness of an estimation amounts to minimizing its
curvature.  The function estimate $\tilde{f}(x)$ is then determined by the
usual least-squares minimization problem with the additional smoothness
constraint.  The amount of smoothing is controlled by the smoothing parameter
$\lambda$, which enters the minimization problem:
\begin{equation}
\chi^2=\sum_i\{y_i-\tilde{f}(x_i)\}^2+\lambda \int_a^b
\{\tilde{f}''(x)\}^2 \d x \stackrel{!}{=}\textrm{min.}
\label{eq:minimizing_problem}
\end{equation}
For simplicity, we have assumed equal measurement uncertainty for all data
points $\sigma_i=const$ and neglected it in the formulas.  The first term
measures the least squares error of the fit, while the second term penalizes
curvature in the function. Equation (\ref{eq:minimizing_problem}) is a a
typical bias-variance problem \cite{Honerkamp-94}, and the best choice of the
smoothing parameter is nontrivial. Numerically, it can be determined using
generalized cross-validation \cite{Gu-Wahba-91}.

As mentioned above, $\tilde{f}$ is usually represented by a superposition of
some basis functions with according coefficients, $a_j$. This is inserted into
(\ref{eq:minimizing_problem}).  The minimizing coefficients, $a_j$, are
determined by a variational principle.  As a consequence the conditions
$\partial \chi^2 / \partial a_j = 0$ have to be fulfilled and the resulting
set of equations needs to be solved (see App. \ref{sec:app}). If we require
$f\in C^2$ to be twice differentiable, natural cubic splines are an obvious
choice. For periodic functions, or $f\in C^\infty$, a spectral representation
suits well.  In other situations, other basis systems might be favorable. In
the following we consider {\it smoothing splines} and Fourier representation.

We want to hint at this point that smoothing splines are different from the
concept of interpolating splines. This is a very important point for the
understanding of the following. Interpolating splines fit a polynomial through
{\it every} data point - in our case a horrible scenario, since we are faced
with noise. Smoothing splines fit a low-order polynomial through bunches of
data, such that the resulting approximation is best in a least-square sense, cf. Eqs.
(\ref{eq:sm_s}) and (\ref{eq:minimizing_problem}).  Summarizing, we note that
the smoothing and the interpolating splines have the representation as
low-order polynomial of nth order (typically 3rd order) in common.
Differences lie in the construction. Whereas interpolation puts a smooth line
between points, smoothing splines use regression to put the best ``line'' (or a
n-dimensional surface, in general) through many points.  Both, interpolating
or smoothing ones, are global, because there is no jump in the nth-order
spline over the whole definition range - nor in the n-1 possible derivatives!
Furthermore, the smoothing spline is constructed from a global measure
(\ref{eq:sm_s}). This measure couples the local spline coefficients.

In \cite{Hanke-Scherzer-2001}, it has been shown rigorously that the minimizer
of (\ref{eq:minimizing_problem}) is a natural cubic spline, provided the
function $f$ is square integrable.  The spline representation used throughout
this paper reads
\begin{equation}
f(x) =  \sum_{j=0}^{n+2} \gamma_j B_j(x)\textrm{,}
\label{eq:bspline_definition}
\end{equation}
where $\gamma_j$ are the coefficients of the cubic B-spline basis functions
$B_j(x)$ and $n$ is the number of knots for constructing the smoothing spline.
After the solution of the minimization problem one calculates the derivative
analytically from the basis functions. The property, important from a
fundamental point of view is the smoothness of the splines.  As shown in
\cite{Hanke-Scherzer-2001}, $e\sim \sqrt{\delta}$ for $\Delta x \to 0$, which
is superior to the accuracy (\ref{eq:error_fd}) for finite difference methods,
especially in the case of very fine sampling.

In the case of a spectral estimate, one writes
\begin{equation}
\tilde{f}(x) = \sum_{k=-n/2}^{n/2} c_k e^{i2 \pi kx /L}
\end{equation}
to be inserted into (\ref{eq:minimizing_problem}). As a result, a system of
equations is obtained for the coefficients $c_k \in \mathbb{C}$. If $n=N$ and
$\lambda=0$, the data are exactly interpolated (Fourier transformed), for
$n\le N$ and/or $\lambda \neq 0$, a spectral smoothing problem results. This
can be solved (see App. \ref{sec:app}), but we will not consider this case.
Instead of solving the complete smoothing problem in spectral space we follow
a slightly different strategy: many experiments suggest that the noise sits
predominantly in the high frequencies.  Then a low-pass filter can be applied.
A well-behaved standard filter is the Butterworth filter
\cite{Smith-Signal-97}, which reads for the m-th\ order:
\begin{equation}
\label{eq:Butterworth}
B(k,k_0)=\frac{1}{1+\Big(\frac{k}{k_0}\Big)^{2 m}}\textrm{,}
\end{equation}
where $k$ is the frequency, $k_0$ is the cutoff-frequency and $m$ defines the
steepness of the filter. The two latter parameters must be related to the
noise in the data. In the case of white noise, $k_0$ can be chosen just above
the last $k$ which occurs from the dynamics of the system, the steepness
determines how ``fast'' the noise amplitude is damped away spectrally after
the cutoff frequency, this can be varied in each case, typical values lie
between 6 and 10 \cite{NR}. So one simply performs a Fourier transformation,
applies the  spectral filter and transforms back. The
derivative is then obtained in spectral space by multiplication with $(i 2\pi k)/
L$ (remember that $y'(x) = \sum_k \frac{i 2 \pi k}{L} c_k e^{i2\pi kx/L}$).
The complete representation of $\tilde{f}$ is
\begin{equation}
\tilde{f}(x)=\sum_{k=-N/2}^{k=N/2} c_k B(k,k_0) e^{i2 \pi kx /L}
=\sum_{k=-N/2}^{k=N/2} \hat{c_k} e^{i2 \pi kx /L} \textrm{,}
\end{equation}
where $c_k$ are the coefficients obtained by Fourier transformation and
$\hat{c_k}=c_k B(k,k_0)$.  It should be mentioned here that this
representation also leads to a smoothing problem
(\ref{eq:minimizing_problem}), cf. App. \ref{sec:app}. But instead of the
computation of a complete set of coefficients $a_j$ one determines the cut-off
frequency $k_0$ optimal for a given $\lambda$, since $k_0$ is the only
parameter to be varied. This means that $\lambda$ and $k_0$ are equivalent and
directly related as is shown in Appendix \ref{sec:app}.  As an alternative to the
Butterworth filter, one can use the Wiener filter, which is a good choice for
linear processes. For scaling systems \cite{BoJePaVu-98}, wavelet transforms
\cite{Holschneider-95,Aldroubi-Unser-96} are a preferred choice.  This
underlines one obvious feature: the more information about the underlying
process is known, the more details one uses for constructing the best method to
estimate derivatives.

Both ways to solve Eq.~(\ref{eq:minimizing_problem}), spline and spectral
method, yield $n$ equations for $n$ unknowns ($n$ being the number of knots
for the smoothing splines or the number of basis function in the spectral
method).  The equations are overdetermined, since $N\ge n$ data points are
available.  This is resolved by the sum in the minimization procedure which
eventually yields an $n \times n$ matrix. The procedure is similar to the
usual linear regression \cite{NR,Hastie-Tibshirani-90}. If a Butterworth
filter is applied, the matrices are reduced to dimension 1.

\section{Numerical Results}
\label{sec:results}
We compare the above methods by three examples, two using numerical data, one
using experimental data: 1) the sine function, 2) the Lorenz system
\cite{Lorenz-63} in a chaotic state and 3) data from an acoustic measurement
\cite{Abel-Bergweiler-Multhaupt-06}.  We quantify our results with the mean
square error of the estimate of the first derivative and its smoothness (or
equivalently its curvature). The dependence on the parameter of the methods
are discussed in detail.

The variance
\begin{equation}
\Delta=<(\tilde{f'}(x_i)-f'(x_i))^2>
\label{eq:mean_square_error}
\end{equation}
is given as an overall measure for the results. The derivative is known exactly
for the numerical data. As a second measure we use the curvature $s(f)$ from
(\ref{eq:sm_s}) or the difference
\begin{equation}
S=\left(s(f)-s(\tilde{f})\right)^2=\left(\int_a^b|\tilde{f}''(x)|^2dx-\int_a^b|f''(x_i)|^2dx\right)^2\textrm{.}
\label{eq:smoothness_difference}
\end{equation}
Other measures can be used, e.g., correlations or a norm different from $L_2$.
We would like to point out that the characterization of results by
(\ref{eq:mean_square_error}) is common to local and global methods, but the
latter use as additional constraint the smoothness. One of the main concerns
of this publication is hint to the importance, from physical arguments and
mathematical considerations, of this property.

For the finite differences we choose a second order method, where the
parameter to be varied is the width $\delta x$. The approximation of
the derivative reads
\begin{equation}
\Big(\frac{\partial \tilde{f}}{\partial x}\Big)_i = \frac{4}{3}
  \frac{y_{i+l}-y_{i-l}}{2 \delta x} -
  \frac{1}{3}\frac{y_{i+2l}-y_{i-2l}}{4 \delta x}\textrm{,}
\label{eq:FD}
\end{equation}
where $l$ is a positive integer determining the step width and $\delta x=l
\Delta x$. The Savitzky-Golay filter was of fourth order; parameter dependence
on the window size $n_l=n_r=n$ has been investigated. The spectral estimator
has as parameters the cut-off $k_0$ and the number of basis functions, which
is set here to $n=N$. The splines have the number of knots, $n$, used as
parameter. The knots are, for the ease of use, space equidistant. Additionally,
for splines and spectral method, one has the smoothing parameter to be varied.
In the spectral case the variation of $\lambda$ is equivalent to the variation
of $k_0$. According to the above, we approximate the function and then
determine the derivative by one of the methods under consideration.

A comparison of the used methods requires a scaling of the parameters. For the
finite difference method we use the quantity $w_{FD}=4 l \Delta x=4 \delta x$
the distance between the leftmost and the rightmost point in the considered
domain. For Savitzky-Golay filter we choose the window size $w_{SG}=(2
n+1)\Delta x$. $w_{SG}$ which corresponds directly to $w_{FD}$. For the
spectral method, we define the window size by the wavelength of the cut-off,
$w_S=L/ k_0$ (since $n=N$). For the smoothing splines, finally, the typical
window can be described by $w_{SM}=L/n$, the distance between two knots.
\begin{figure}
  \begin{center}
    \includegraphics[draft=false,angle=270,width=0.7\textwidth]{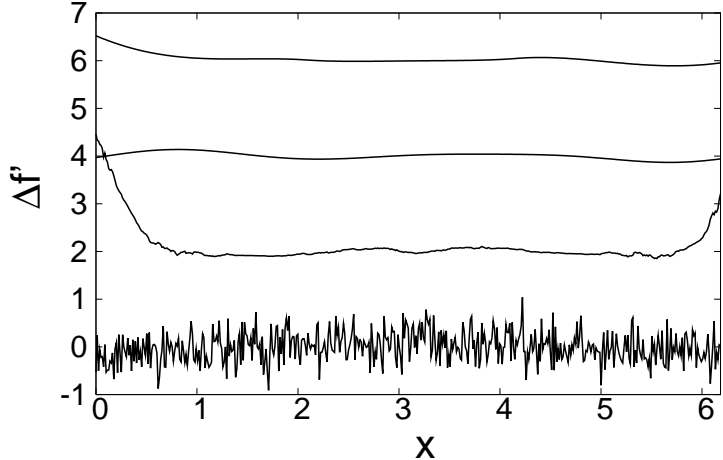}
  \end{center}
  \caption{Difference $\Delta f'(x_i)=\tilde{f}'(x_i)-f'(x_i)$ of the
    estimate and the exact derivative $f'(x)=\cos(x)$. Measurement noise was
    simulated by a Gaussian white noise process with standard deviation
    $\sigma=0.5$. From bottom to top: 1. second order finite difference
    method ($\delta x=1.55$, $w_{FD}=6.18$) 2.  Savitzky-Golay-filter
    ($n=143$, $w_{SG}=3.61$), 3. spectral method ($k_0=2.99$, $w_S=2.09$) and
    4. smoothing splines ($n=7$, $\lambda=0.13$, $w_{SM}=0.89$). Parameters
    for the methods were chosen such that they were optimal in the sense of
    Eq.~(\ref{eq:mean_square_error}). The offsets 0, 2, 4, 6 are added
    respectively for better visibility.}
  \label{fig:derivs_sine}
\end{figure}

\subsection{Sine Function}
\label{sec:result:sine}
As a  first example we use the function $f(x)=\sin (x)$ defined on the
interval $0 \le x \le 2\pi$ with added Gaussian white noise $\eta_i$
with standard deviation $\sigma$ and zero-mean. The data set consists of 
500 Points $y_i=f(x_i)+\eta_i$, so that $\Delta x=2\pi / 500$. In
Fig.~\ref{fig:derivs_sine} the result for the derivative estimate is
displayed ($\sigma=0.5$). Finite differences ($\delta x=1.55$,
$w_{FD}=6.18$) yield an unacceptable result with extreme fluctuations
of the order of $0.1$. Also, the result for the Savitzky-Golay filter
($n=143$, $w_{SG}=3.61$) is not smooth and deviates heavily at the
boundaries. Spectral ($k_0=2.99$, $w_S=2.09$) and spline ($n=7$,
$\lambda=0.13$, $w_{SM}=0.89$) methods, apparently work better.
\begin{figure}
  \begin{center}
    \subfigure[]{
      \includegraphics[draft=false,angle=270,width=0.4\textwidth]{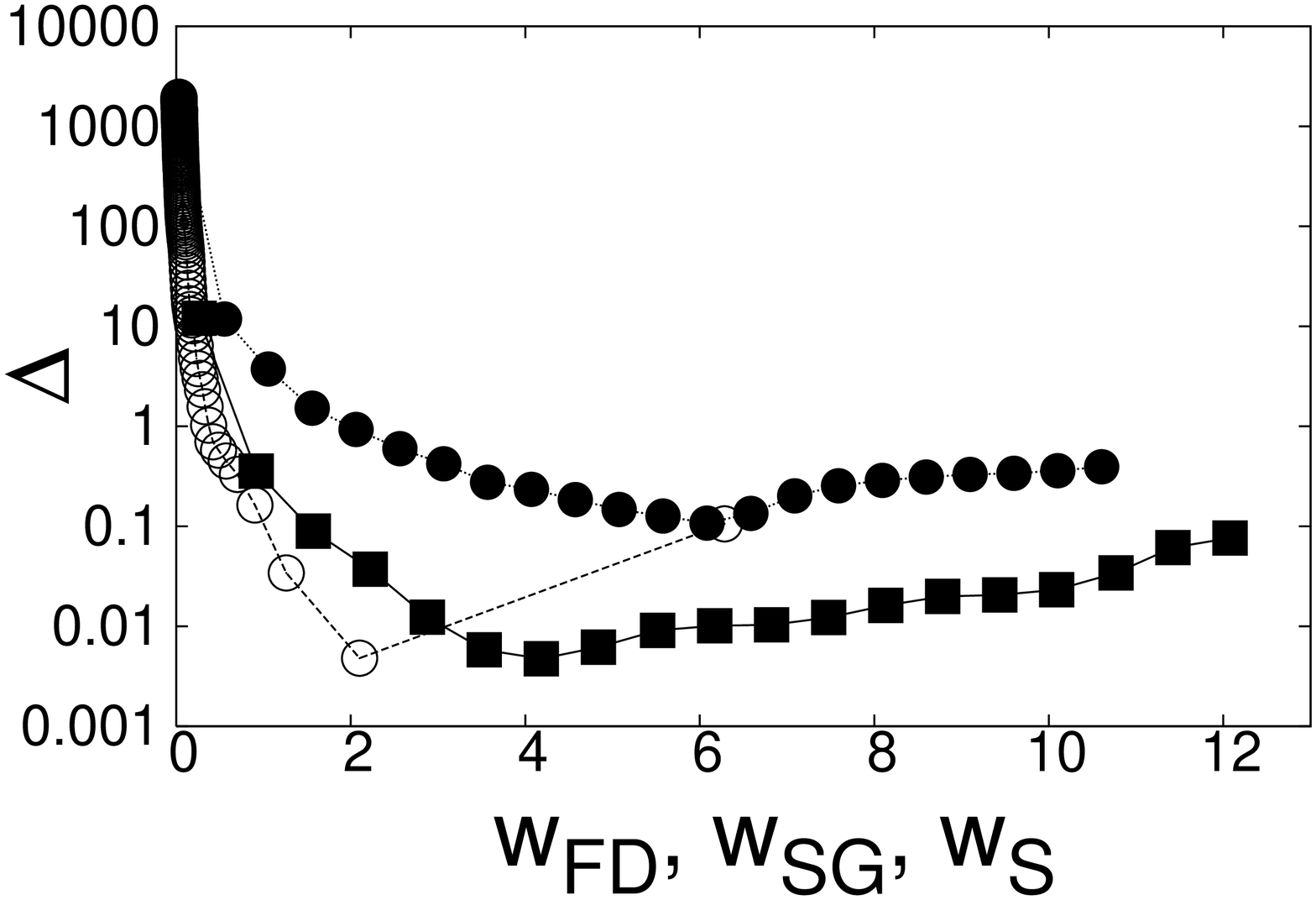}
    }
    \subfigure[]{
      \includegraphics[draft=false,angle=270,width=0.4\textwidth]{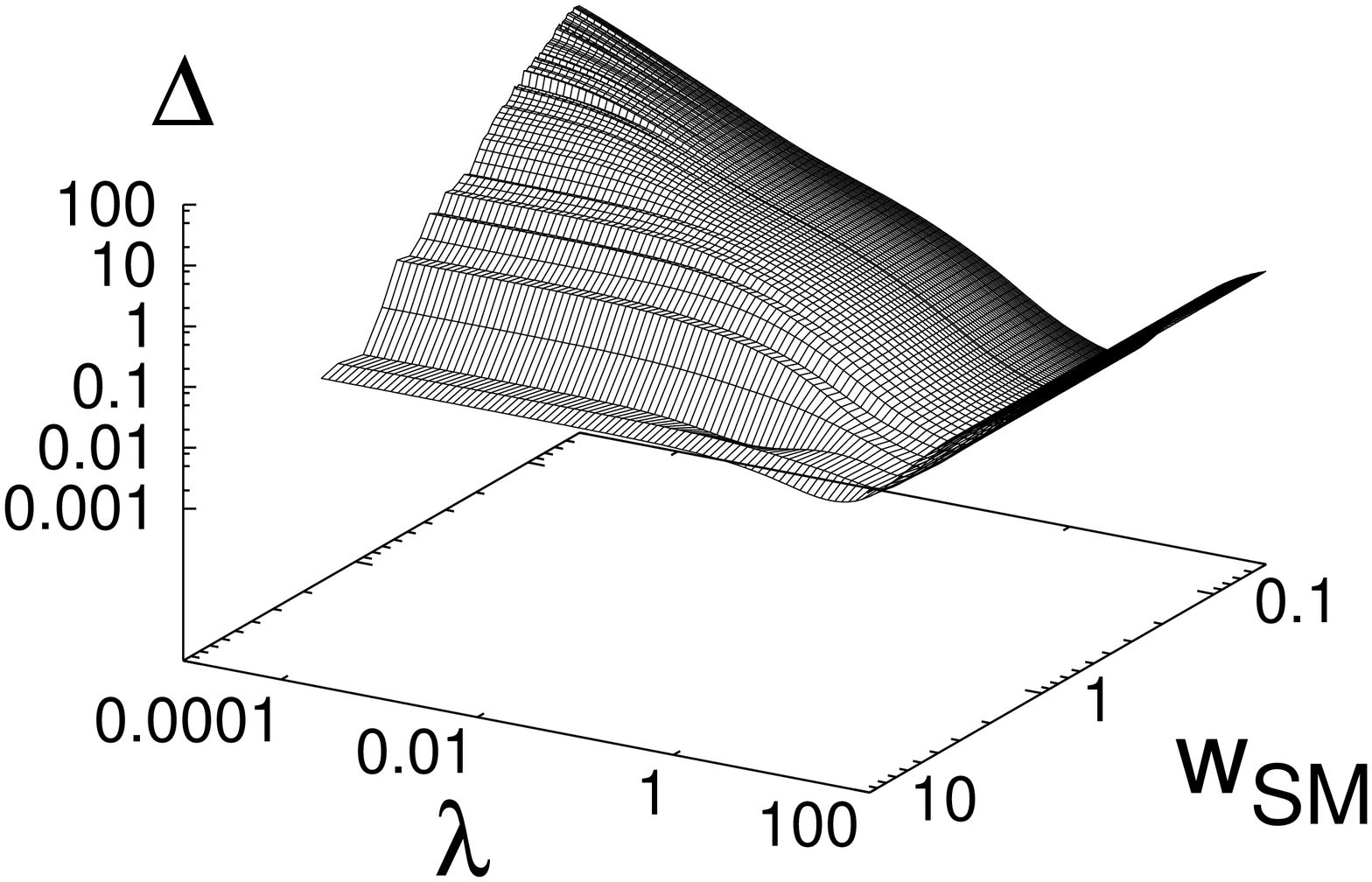}
    }  
  \end{center}
  \caption{Parameter dependence of the mean square error of the derivative of
    $\sin (x)$ with additive with noise $\sigma=0.5$. (a) For the finite
    differences, Savitzky-Golay-filter and the spectral method. $\bullet$ --
    finite difference method, the according parameter is $w_{FD}$,
    $\blacksquare$ -- Savitzky-Golay-filter ($w_{SG}$), $\circ$ -- spectral
    method ($w_S$). (b) mean square error for the smoothing splines in
    dependency of $w_{SM}$ and $\lambda$.}
  \label{fig:error_sine}
\end{figure}

Up to now, we showed the result for a specific set of parameters for
each method. For a complete analysis, the dependence of the different
methods on their parameters needs to be investigated.  We varied the
parameters over a wide range and considered the mean square error
(\ref{eq:mean_square_error}), cf. Fig.~\ref{fig:error_sine}. For every
method a minimum occurs at the point of  optimal approximation in the
least-squares sense.

For the finite difference method we varied the width $\delta x$. The optimal
width is $\delta x=1.55$ ($w_{FD}=6.18$), nearly $1/3$ of the domain. Smaller
spacing results in larger fluctuations, whereas an increasing spacing will not
approximate the desired derivative. It is clearly visible from
Fig.~\ref{fig:derivs_sine} that finite differences show the strongest
fluctuations among all considered techniques. In practice finite differences
should not be the first choice.

For the Savitzky-Golay-filter we varied the window size to find an optimum at
$n\approx143$ ($w_{SG}=3.61$). For a larger window the filter smoothes the
function too much, and the result tends to a constant.  For a smaller window
the influence of noise becomes locally more important.

For spectral differentiation, the filter cut-off $k_0$ can be varied from
$0$-$250$ to determine the optimal cut-off. We find $k_0 \approx 3$
($w_S=2.09$). The original function, $\sin (x)$ implies that only $k=1$ is
active in spectral space; for a top-hat filter with sharp edges this would
result in $k_0=1$. Because $B(k,k_0)$ is smooth, a slightly larger cut-off is
found. For higher cut-off values the approximation increasingly oscillates
around the optimal solution. For spectral differentiation, in general,
problems near boundaries occur, if the data are not perfectly periodic. In
this case some data points close to the boundaries should be discarded after
the determination of the derivative, but it might be better to switch to
splines or other basis functions.

For smoothing splines, the mean square error depends on the number of used
knots and the smoothing parameter. The optimal values are $\lambda\approx
0.13$ and $k=7$ ($w_{SM}=0.89$). When the smoothing parameter is increased the
estimate consistently tends to a constant, when it is decreased the estimate
represents a bigger part of the noisy fluctuations. The meaning of the number
of used knots for the smoothing splines can be understood as the degree of
freedom of the smoother; if there is, e.g., one oscillation in the data, one
needs minimal $3$ knots for approximation. More oscillations require more
knots, corresponding to a higher resolution, e.g., 10 oscillations can not be
resolved with a smoothing spline of 10 knots; for details see
\cite{Wahba-90}. In this sense the degree of freedom of the smoother can be
understood as ability to fit a given number of oscillations in the data. If
the number of knots exceeds the number of oscillations, one is faced with the
bias-variance problem that all methods have in common.
\begin{figure}
  \begin{center}
    \subfigure[]{
      \includegraphics[draft=false,angle=270,width=0.4\textwidth]{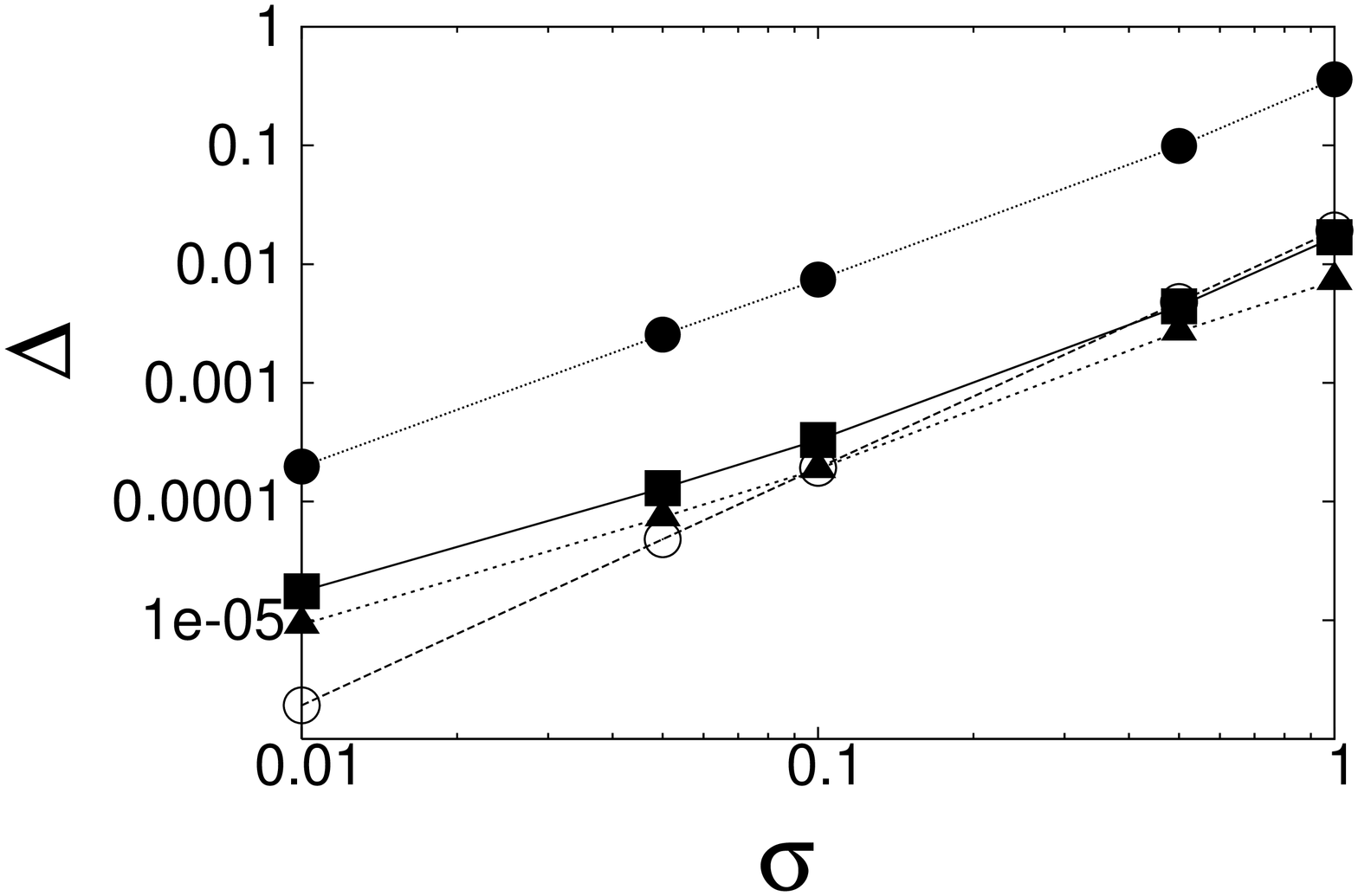}
    }
    \subfigure[]{
      \includegraphics[draft=false,angle=270,width=0.4\textwidth]{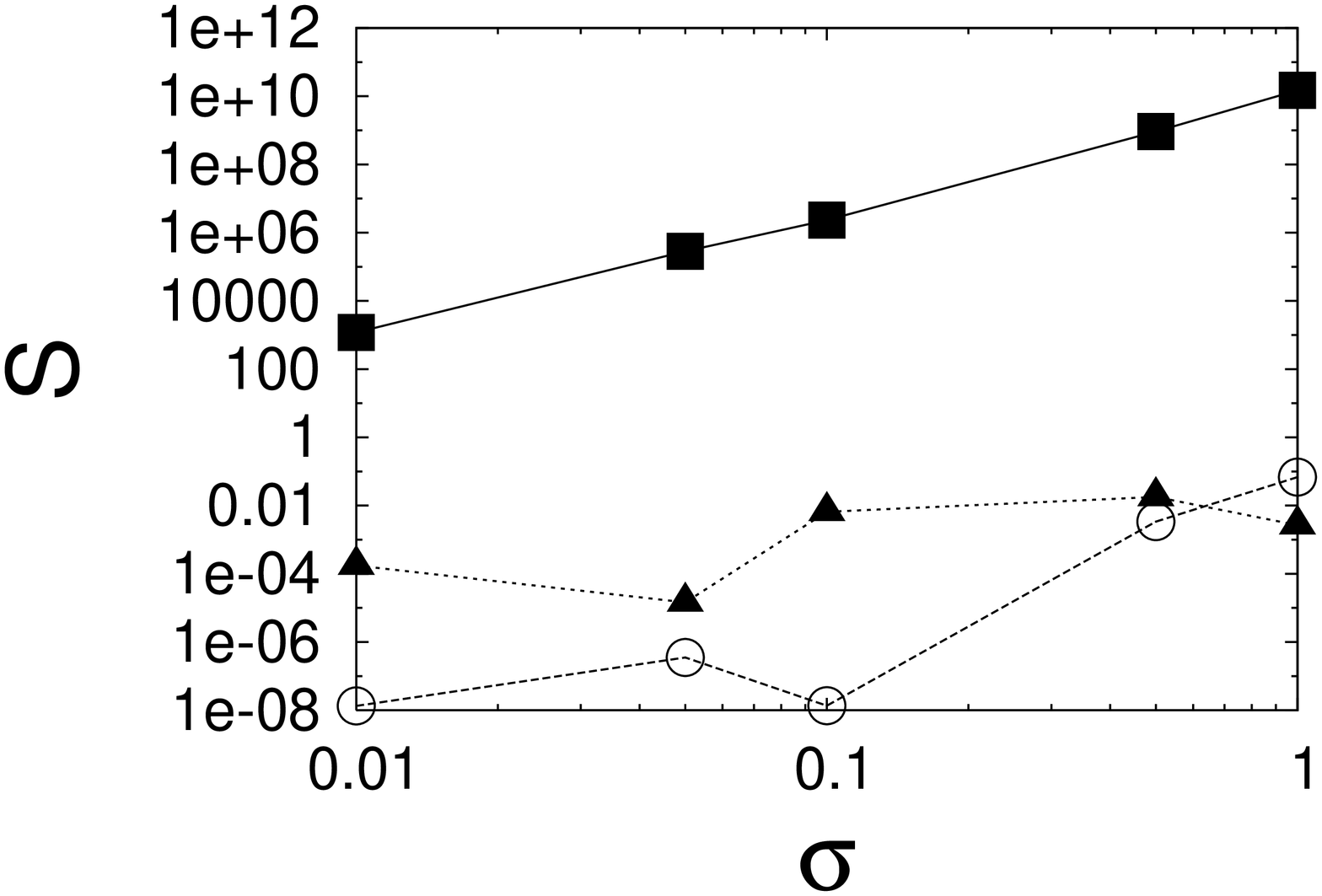}
    }
  \end{center}
  \caption{Characterization of the derivative estimate of the sine data for
    different noise levels. a) least squares error, b) curvature $S$. On the
    x-axis the noise level of the white Gaussian noise added to the signal is
    plotted.  $\bullet$ -- finite difference method, $\blacksquare$ --
    Savitzky-Golay-filter, $\blacktriangle$ -- smoothing splines, $\circ$ --
    spectral method.}
  \label{fig:sine}
\end{figure}

For a direct comparison of the investigated methods for numerical
differentiation we studied the dependence of the mean square error
(\ref{eq:mean_square_error}) on the noise level $\sigma$ for each method.
Results are shown in Fig. \ref{fig:sine}a. Finite differences have an
error about one order of magnitude larger than the other methods. The spectral
method works very well for a small noise level, with an error of one order
smaller than the Savitzky-Golay-filter and smoothing splines. For a higher
noise contamination, however, smoothing splines are the best choice, whereas
spectral methods and the Savitzky-Golay-filter are comparable by means of
(\ref{eq:mean_square_error}).  Nevertheless, from a mathematical point of
view, the Savitzky-Golay method is a local approximation, with no smooth
relation between $\tilde{y}'(x_i)$ and $\tilde{y}'(x_{i+1})$.  This is also
visible in Fig.\ \ref{fig:derivs_sine} where at several points jumps can be
observed.

To quantify the intuition on smoothness from the graphs
(Fig.~\ref{fig:derivs_sine}), we studied the dependence of the curvature $S$
on the standard deviation of the noise, cf.
Eq.~(\ref{eq:smoothness_difference}). The estimate was interpolated by a
spline and then $S(\tilde{f})$ has been computed from (\ref{eq:sm_s}) by
averaging $|\tilde{f}''(x)|^2$ over each interpolated data point. The
interpolation is necessary to determine the second derivative $\tilde{f}''$.
One obtains $s(f)$ analytically, so that $S$ is easily determined. The results
are shown in Fig.~\ref{fig:sine}b. The curvature of the Savitzky-Golay-filter
is some orders of magnitude worse than the one for the spectral method and the
smoothing splines.  Finite difference techniques are  worse than
Savitzky-Golay-filter, so that we did not consider it here.  The
main difference between the methods is the different behavior of the
smoothness, or the curvature, respectively. While for the Savitzky-Golay-filter one can
clearly see a linear relation between noise and smoothness, for the spectral
method and the splines the relation is roughly a constant, up to a certain,
parameter-dependent noise level. One recognizes large fluctuations of $S$,
resulting from over-smoothing the fit.

\subsection{Lorenz System}
\label{sec:result:lorenz}

\begin{figure}
  \begin{center}
    \includegraphics[draft=false,angle=270,width=0.8\textwidth]{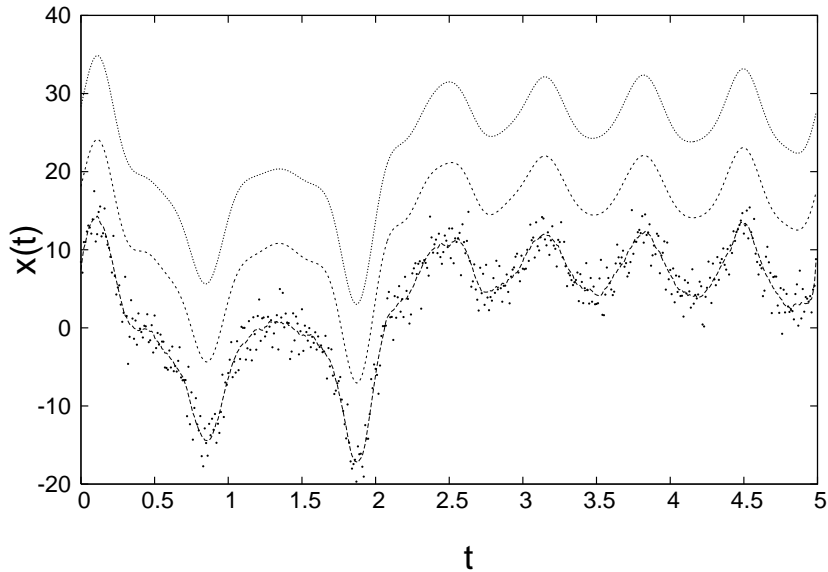}
  \end{center}
  \caption{Time dependency of the $x$-component of the Lorenz system, with
    additive with noise $\sigma=2.0$. Also shown are the results of the
    regression: Savitzky-Golay filter with $w_{SG}=0.47$ (bottom), spectral
    method $w_{S}=0.25$ (middle) and smoothing splines $w_{SM}=0.13$
    (top). An offset is added for better visibility.}
  \label{fig:lorenz_trajec}
\end{figure}

As a second example we analyzed the $x$-component of the Lorenz system
(Fig.~\ref{fig:lorenz_trajec}),
\begin{subequations}
\label{eq:lorenz}
\begin{eqnarray}
\dot{x} & = & \sigma(y-x) \label{eq:lorenz_x} \\
\dot{y} & = & R x-y-xz    \label{eq:lorenz_y} \\
\dot{z} & = & -bz +xy \textrm{.} \label{eq:lorenz_z}
\end{eqnarray}
\end{subequations}
We integrate the system numerically by a Runge-Kutta algorithm of fourth order
with time-step 0.01 and parameters $\sigma=10$, $R=28$, $b=8/3$. The
integration was performed over 500 steps only. We also added Gaussian white
noise to the data. The $x$ component of the trajectory of (\ref{eq:lorenz}) is
shown in Fig.~\ref{fig:lorenz_trajec}. Results for the dependence of the mean
square error on the level of noise are similar to the previous case,
$f(x)=\sin (x)$ (see Fig.\ \ref{fig:lorenz}). For small noise the spectral
method provides the smallest approximation error. Finite differences are here
of the same order of magnitude.

\begin{figure}
  \begin{center}
    \subfigure[]{
      \includegraphics[draft=false,angle=270,width=0.4\textwidth]{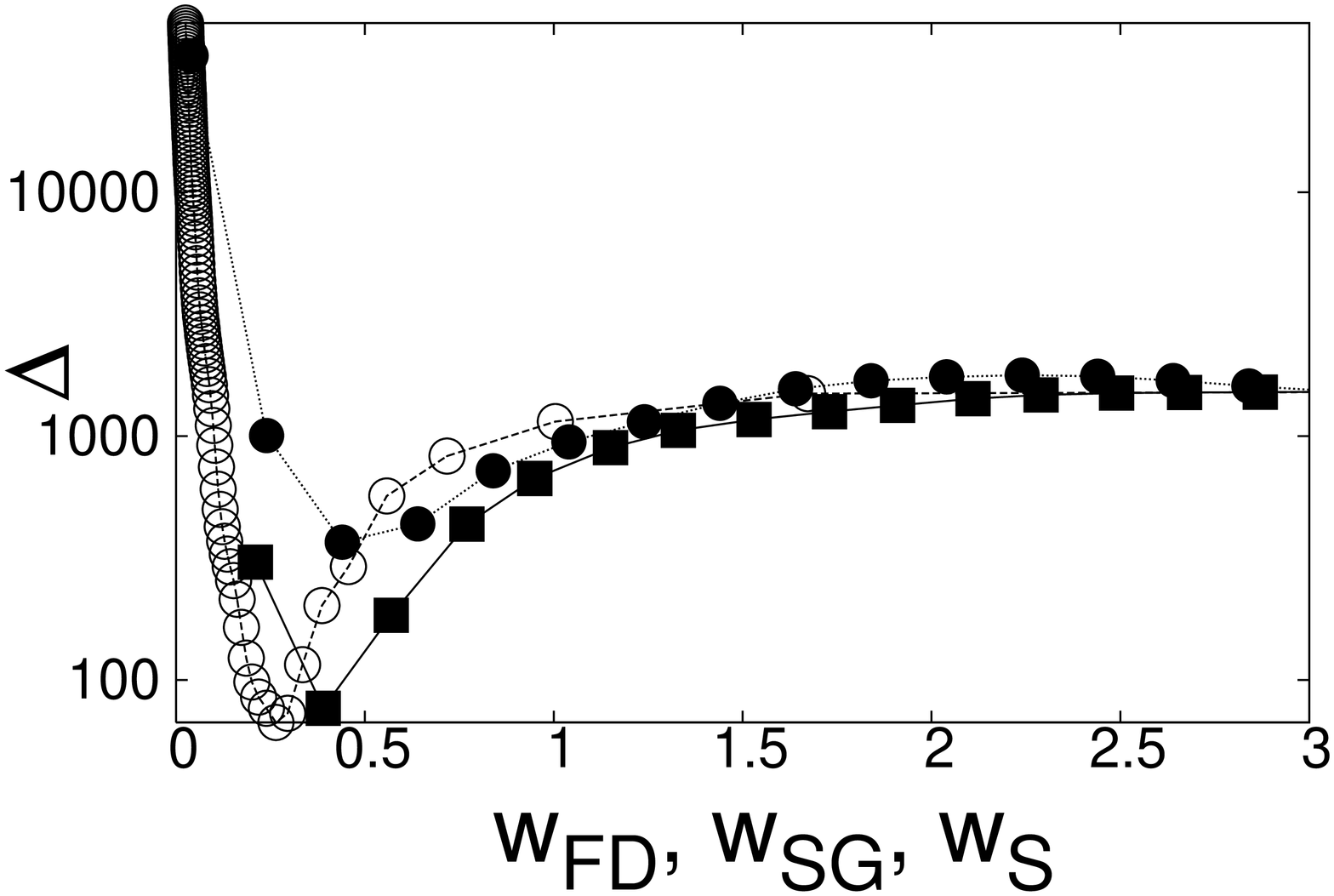}
    }
    \subfigure[]{
      \includegraphics[draft=false,angle=270,width=0.4\textwidth]{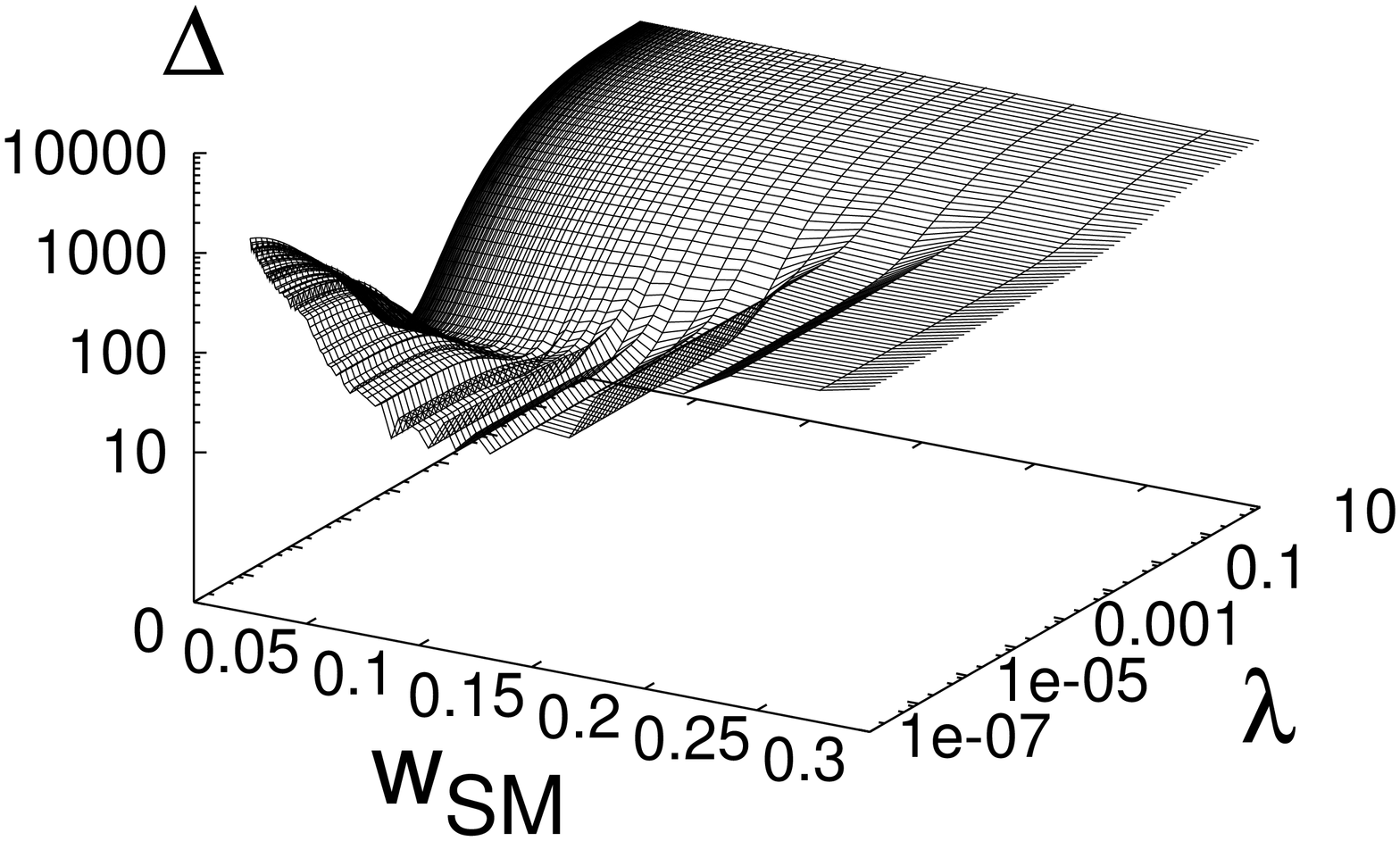}
    }
  \end{center}
  \caption{Parameter dependence of the least squares error for the
    $x$-component of the Lorenz system with additive white noise $\sigma=2.0$
    (a) For the finite differences, Savitzky-Golay-filter and the spectral
    method.  $\bullet$ -- finite difference method, the according parameter is
    $w_{FD}$, $\blacksquare$ -- Savitzky-Golay-filter ($w_{SG}$), $\circ$ --
    spectral method ($w_S$). (b) least squares error for the smoothing splines
    in dependence on the number of knots $k$ and the smoothing parameter
    $\lambda$.}
  \label{fig:error_lorenz}
\end{figure}

\begin{figure}
  \begin{center}
    \subfigure[]{
      \includegraphics[draft=false,angle=270,width=0.4\textwidth]{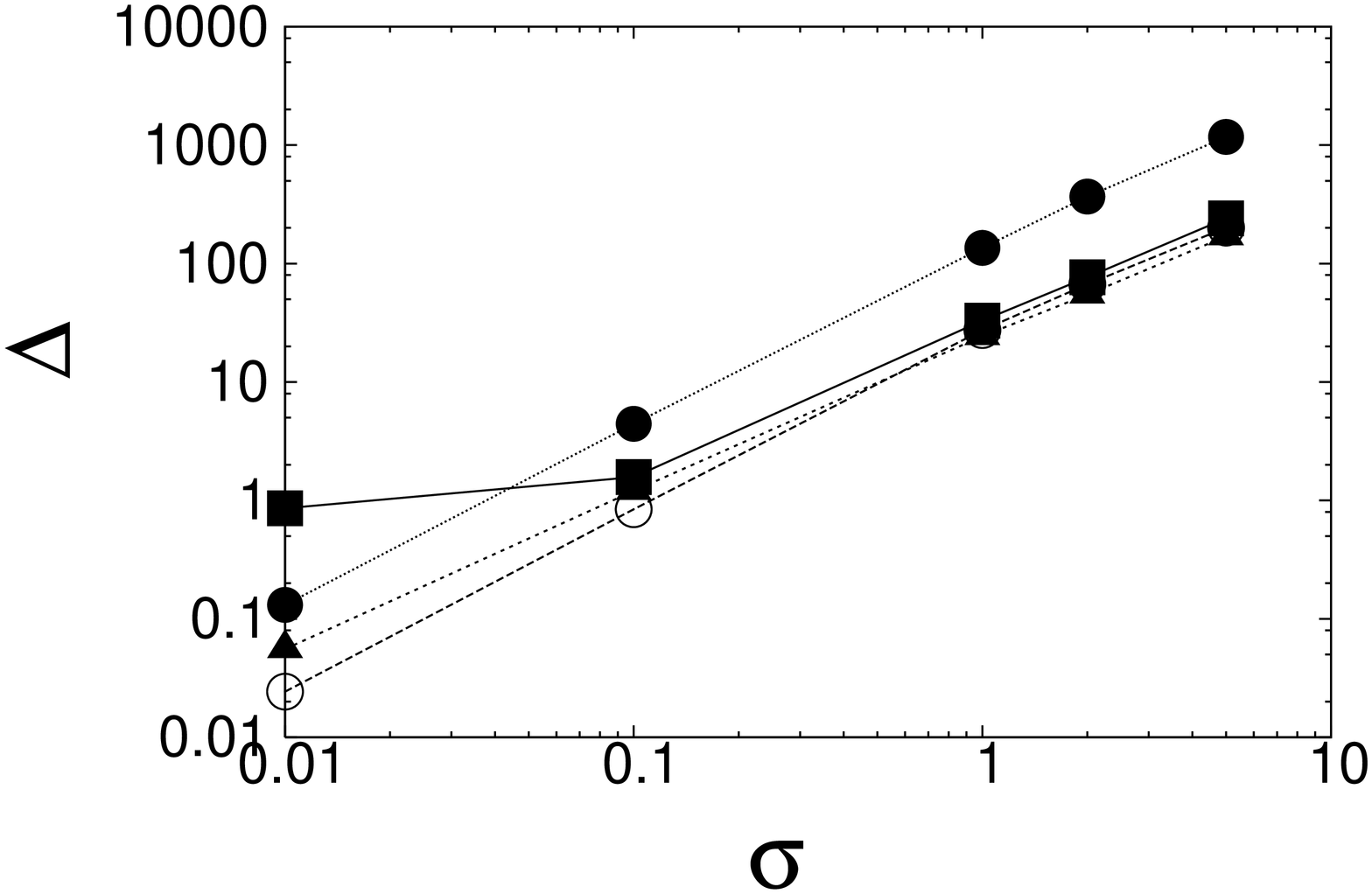}
    }
    \subfigure[]{
      \includegraphics[draft=false,angle=270,width=0.4\textwidth]{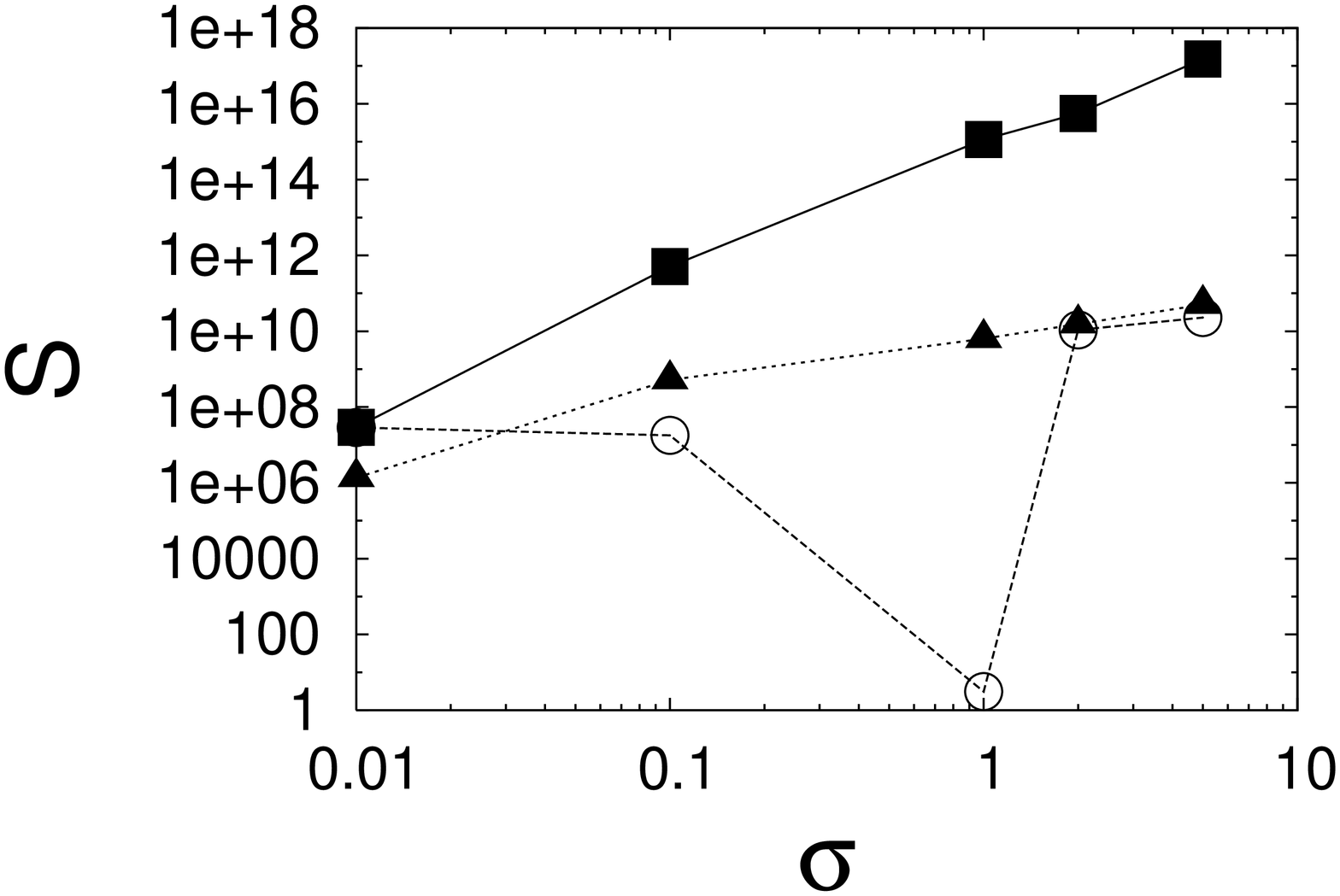}
    }
  \end{center}
  \caption{Dependence of the least squares error and the smoothness on the
    additive noise standard deviation for the $x$-component of the Lorenz
    system, Eq.~ (\ref{eq:lorenz}).  a) least squares error, b) curvature
    $S$. On the x-axis the noise level of the white Gaussian noise added to
    the signal is plotted.  $\blacksquare$ -- Savitzky-Golay-filter,
    $\blacktriangle$ -- smoothing splines, $\circ$ -- spectral method. The
    parameters for the fit are chosen to be optimal, cf.
    Fig.~\ref{fig:error_lorenz}.}
  \label{fig:lorenz}
\end{figure}

Because in a chaotic time series, many scales are present with a broad
spectrum, the Lorenz system is a good candidate for a study of the
parameter dependence of the estimates under the aspect of a scaling
system. We show the comparison of the methods in Fig.
\ref{fig:error_lorenz}.  The optimal parameters found here are
different from the values for the sine. This is explained by the
different number of oscillations. The data set for $\sin (x)$ contains
exactly one oscillation, whereas the data for the Lorenz system
contains 6 oscillations.  As mentioned above the parameters must be
chosen  such that  the bias-variance trade-off  encountered.

For the Savitzky-Golay-filter, $w_{SG}=3.61$ for $\sin (x)$, whereas
$w_{SG}=0.47$ for the Lorenz system. This difference can be explained by the
fact that large window sizes will use many points for fitting a polynomial of
4th order to each data point. For data sets with many oscillations this will
result in bad approximation of the derivative or the function. For the
smoothing splines $w_{SM}=0.89$ for $\sin (x)$, whereas for the Lorenz system
$w_{SM}=0.13$. To resolve many oscillations, obviously more knots are
needed. In the spectral case the dependency of the cut-off is $w_{S}=2.09$ for
$\sin (x)$ and $w_{S}=0.26$ for the Lorenz system.  Summarizing this
subsection, the results for the much more complicated chaotic time signal
match the ones for the quite simple sine signal.  This can be interpreted as a
sign for the generality of the results.

For a comparison of numerical differentiation methods with different data
series one usually rescales the $x$ axis (in case of the Lorenz system
the time $t$) by $\tilde{x}=x N_{peak}/L$, where $N_{peak}$ is the number of
peaks. Then, the regression parameters $w_{FD}$, $w_{SG}$, $w_{SM}$ and $w_S$
are of the same order. But furthermore it should be noted, that every data
series possesses characteristic edges, which will naturally result in
different parameters. If the time series also consists of intermittent
domains, the situation is even worse. Then $N_{peak}/L$ is not a measure
for the mean 'wavelength' of the data, and the domains, containing the
oscillations would be over-smoothed. A detailed discussion of such phenomena
goes far beyond the scope of this article.

\subsection{Experimental Data}
\label{sec:result:chemical}
As the last example we analyzed experimental data from the acoustical signal
emitted from the mouth of an organ pipe. This measurement is needed, if the
complex acoustical system of an organ pipe shall be modeled as a nonlinear
oscillator \cite{Fabre-00} to be found numerically by nonparametric data
analysis \cite{Voss-Buenner-Abel-98,Abel-04}. Basically, the physics of the
measured data it is not very important for our purposes and we will not
comment further on the origin of our data, details are found elsewhere
\cite{Abel-Bergweiler-Multhaupt-06}. The time series $y(t)$ consists again of
500 points, the sampling interval is $\Delta t=1/44100\; s$. In contrast to the
previous examples, we do not have separate access to the derivatives and the
measures (\ref{eq:mean_square_error}) and (\ref{eq:smoothness_difference}) can
not be applied to determined the regression parameters. So, one needs to
estimate the optimal parameters. In principle, there are methods like {\it
  generalized cross-validation} dealing with this problem
\cite{Hastie-Tibshirani-90,Gu-Wahba-91,Davison-03,Green-Silverman-94}. We will
now explain briefly the functioning of generalized cross-validation (GCV) to
have a complete presentation, an explicit application to our data lies beyond
the scope of this article and is subject to future work.
\begin{figure}
  \begin{center}
    \includegraphics[draft=false,angle=270,width=0.8\textwidth]{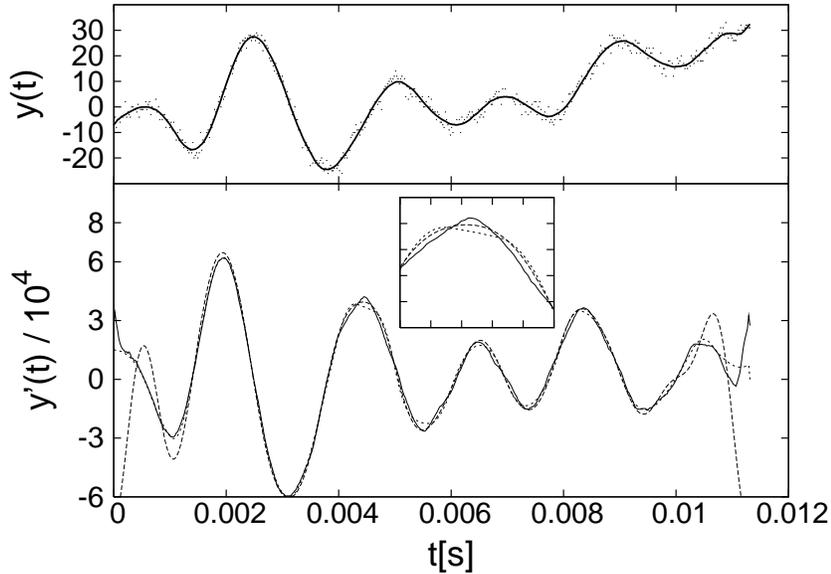}
  \end{center}
  \caption{Upper plot: data points of the organ pipe measurements and fitted
  function using Savitzky-Golay filter. The other methods are
  indistinguishable by eye. Lower plot: fitted derivatives. Solid line:
  Savitzky-Golay filter, dashed line: spectral filtering, dotted line:
  smoothing splines. The inset shows a part of the fitted derivative in the
  interval $0.04 s \le t \le 0.05s$ Without quantitative analysis it is hard
  to say which curve is best, differences are however recognizable. The
  parameter used for fitting are $w_{SG}=1.6 \cdot 10^{-3}s$ for
  Savitzky-Golay filter, $w_{S}=8.7 \cdot 10^{-4}s$ for the spectral method
  and $w_{SM}=3.2 \cdot 10^{-4}$ and $\lambda=2.6\cdot 10^{-11}$ for the
  smoothing splines.}
  \label{fig:pipe_fit}
\end{figure}

GCV works by dropping one point $(y_i,x_i)$ from the data (in general $M$
points), the estimate is then based on the remaining $N-1$ points. The GCV is
constructed by sum of squares
\begin{equation}
GCV(\lambda)=\frac{1}{N}\sum_{i=1}^{N}\{y_i-\tilde{f}_\lambda^{-i}(x_i)\}^2
\textrm{.}
\label{eq:cv}
\end{equation}
$\tilde{f}_\lambda^{-i}$ is the estimate for $f(x)$ if the point $(y_i,x_i)$
is omitted. The optimal parameter is calculated from $GCV$ for a number of
values of $\lambda$ over a suitable range, then the minimizing $\lambda$ is
selected.  In other words, Eq.~(\ref{eq:cv}) gives an empirical estimate of
the combined systematic errors, induced by the choice of the parameters, and
statistical errors due to noise in the data. The parameter is chosen
to minimize that estimate. Note that this is done for the estimated function,
not the derivative.

Results for the estimates of function and derivative of the experimental data
are shown in Fig. \ref{fig:pipe_fit}. For the experimental data we are faced
with the problem that we do not know the true derivative. So, we have to
compare the estimation, or construction of the function itself in terms of
least-squares error $e=<(\tilde{f}(x)-f(x))^2>$ and the corresponding
curvature $s$.  All methods approximate the function quite well in terms of
the least-squares criterion.  As well the derivative seems well estimated, but
slight differences are recognized. First, one notices the weakness of the
spectral estimator at the boundaries; as well splines and Savitzky-Golay
filter should be considered with great care in this region, as we already saw
in the previous sections. By eye, all three estimates appear
indistinguishable. However, if we estimate the curvature we find great
differences.

We have two competing quantities in our considerations, the least-squares
error, and the curvature.  To compare either of them, one should fix the other
one, correspondingly we need to fix the least-squares error to obtain
quantitative statements about $s$. cf. Eq. (\ref{eq:sm_s}). We -quite arbitrarily-
chose the value $e=2$.  Then, we searched in the whole parameter space of all
the methods for fits with the least-squares error close to 2 and compared the
according curvature.  The Savitzky-Golay technique yields $e=2.04$ and the corresponding
curvature $s=2.05 \cdot 10^{15}$, the window size $w_{SG}=1.61 \cdot 10^{-3}$.
For the spectral estimator, we obtained $e=2.03$ with a curvature of
$s=6.16\cdot 10^{13}$ at $w_{S}=8,76 \cdot 10^{-4}$. The spline estimator yeld
an error $e=2.05$ with $s=5.52\cdot 10^{13}$ and $w_{SM}=3.23\cdot 10^{-4}$.
As a result we find again that the global methods are smoother by two orders
of magnitude in comparison with the Savitzky-Golay method.  Obviously, as far
as a smooth curve is concerned the global smoothers are superior to the local
methods under consideration. If smoothness is not relevant, all three methods are
equivalent if the boundaries are neglected.

\section{Conclusion}
\label{sec:conclusion}
We presented a qualitative and quantitative comparison of local and global
methods for the numerical estimation of derivatives. Whereas local methods are
appealing due to their simplicity and easy implementation, we advocate for
global methods, because the properties of the functions can be defined in a
proper way. We focused on the important constraint of smoothness (expressed by
the curvature) and showed how a corresponding minimization problem is
solved. Furthermore, we demonstrated that global methods are superior to local
ones if high-precision estimates are needed or measurement noise is large. We
did not want to consider in detail the computational cost. But it shall be
mentioned that a global estimate can be expensive if the number of points
exceeds $10^5$. Then, programming skill (or large memory) is required to
encounter the problem of large matrices to be multiplied.

We compared in this article finite differences, Savitzky-Golay filtering,
smoothing splines and smoothing spectral estimators.  To compare the methods,
the dependence on parameters has been investigated, optimal parameters could
be determined. The dependence on additive noise has been studied in detail and
we found enormous differences in the methods.  One result is that finite
differences are orders of magnitudes off in comparison to the other
methods. It is used as a worst case demonstration in this article.  In terms
of the least-squares error the remaining three methods are comparable in the
sense that they are of the same order of magnitude. However in terms of
smoothness, the Savitzky-Golay filter fails by some orders of magnitude, and
only the global methods work well.  It is remarkable that we can compare the
methods on a logarithmic scale, i.e., techniques differ really to a huge
extent.  A more detailed look shows that spectral estimators work very well
for small noise levels. For high noise levels, smoothing splines yield better
results for the investigated systems.

Under a more general perspective, we showed how one can choose among some set
of basis functions for the representation of the estimate.  Obviously, for
periodic functions a spectral representation is natural, similarly if one is
interested in higher derivatives. If other boundary conditions are required,
other basis functions might be favorable. Smoothing splines are an optimal
choice if twice differentiability (or smoothness) is required, and no further
information is available. For all basis systems the general procedure
described above applies, formulated as a minimization problem. In principle
one can imagine further constraints like minimal variance of the first
derivative or other criteria. Those conditions can be easily built into the
method as additional Lagrange multiplier.

From a practical point of view one has to decide how important it is to obtain
smooth functions to a reasonable accuracy. For a rough guess, a local filter
might do, for any high-precision analysis the implementation of the
minimization, or smoothing problem does pay off. E.g., if one wants to process
further the obtained derivatives, small differences can yield enormous changes
in the final results. Our interest started with an application in some
reconstruction techniques \cite{Abel-Ahnert-Kurths-05}, where local
methods are by far too inexact. Similar holds for prediction problems where
the integration of functions based on the estimate of the derivatives is
important \cite{Abarbanel-97}.

\section*{Acknowledgments}
We thank M. Rosenblum and M. Hanke-Bourgeois for helpful discussion. M. Abel
and K. Ahnert acknowledge support by the DFG (German research foundation)
(Proj. Nr. AB143/3). We are very grateful to S. Bergweiler for providing the
experimental data.

\appendix
\section{Appendix: Spectral smoother }
\label{sec:app}
In  the following, we assume that $x \in [0,1]$
for the measured data
$(x_n,y_n) ; n \in 1,\dots,N$

The spectral representation of a function reads
\begin{equation}
f(x)=\sum_{k=0}^{N-1} c_k e^{i 2\pi k x}\;.
\end{equation}
and the smoothing term in  Eq.~(\ref{eq:minimizing_problem}) reads
\begin{eqnarray}
\lambda \int_0^1 | f''(x) |^2 dx & = & \lambda \int_0^1 f''(x)
f''(x)^* dx \\ \nonumber
 & = & 16 \lambda \pi^4 \sum_{k,l=0}^{N-1} c_k c_l^* k^2 l^2 \int_0^1
e^{i 2 \pi (k-l) x }dx \\\nonumber
& = & 16 \lambda \pi^4 \sum_{k=0}^{N-1} c_k c_k^* k^4\;,
\end{eqnarray}
where $\int_0^1 e^{i 2 \pi (k-l) x} dx = \delta_{kl}$ is used. To solve the
minimizing problem Eq.~(\ref{eq:minimizing_problem}), we insert the above into
\begin{eqnarray}
\chi^2 &=& \sum_{n=0}^N | y_n - f(x_n) |^2 + \lambda \int_0^1 |f''(x)|^2
dx \stackrel{!}{=}\textrm{min.}\;,\\ \nonumber
       &=& \sum_{n=0}^N \Big( y_n - \sum_{k=0}^{N-1}c_k e^{i 2 \pi k x_n} \Big)\Big(y_n - \sum_{k=0}^{N-1}c_k^* e^{-i 2 \pi k x_n} \Big) + 16 \lambda \pi ^4
\sum_{k=0}^{N-1} c_k c_k^* k^4\;.
\end{eqnarray}
By variation of the coefficients we obtain the conditions
\begin{equation}
\frac{\partial \chi^2}{\partial c_k}   = 0\;\;\;,\;
\frac{\partial \chi^2}{\partial c_k^*} = 0\;.
\end{equation}
This yields the equations
\begin{eqnarray}
0 & = & -\sum_{n=1}^N e^{i 2\pi k x_n} \left( y_n-\sum_{l=0}^{N-1} c_l^* e^{-i
  2\pi l x_n} \right) + 16 \pi^4 \lambda c_k^* k^4 \\
0 & = & -\sum_{n=1}^N e^{-i 2\pi k x_n} \left( y_n-\sum_{l=0}^{N-1} c_l e^{i
  2\pi l x_n} \right) + 16 \pi^4 \lambda c_k k^4
\end{eqnarray}
These are 2N linear equations for the 2N unknowns $\{c_k, c_k^*\}$ which can
be solved by usual algebraic manipulation.

In the case of using the Butterworth filter (\ref{eq:Butterworth}), one
determines the Fourier coefficients $c_k$ in the conventional way \cite{NR}.
Then the filter is applied. The only possible variation is in $k_0$ and a
single equation results. For simplification we write $B(k,k_0)=B_k$, $\partial
B(k,k_0) / \partial k_0 = B'_k$ and $c_k e^{i 2 \pi k x_n}=C_{kn}$ and obtain:
\begin{eqnarray}
\frac{\partial \chi^2}{\partial k_0}   &=& 0\\ \nonumber
 & = & -\sum_{n=1}^N \Big(
\sum_{k=0}^{N-1} B'_k C_{kn} \big(y_n-\sum_{l=0}^{N-1}B_l C_{ln}^* \big) + 
\sum_{k=0}^{N-1} B'_k C_{kn}^* \big(y_n-\sum_{l=0}^{N-1}B_l C_{ln} \big)
 \Big) + \\ & & + 32 \pi^4 \lambda \sum_{k=0}^{N-1} B_k B'_k k^4 c_k c_k^* \\
& = & -\sum_{n=1}^N \Big( \sum_{k=0}^{N-1} B'_k y_n (C_{kn}+C_{kn}^*)  
 -\sum_{k,l=0}^{N-1} B'_k B_l (C_{kn} C_{ln}^* + C_{kn}^* C_{ln}) \Big) \\
& & + 32 \pi^4 \lambda \sum_{k=0}^{N-1} B_k B'_k k^4 c_k c_k^*
\end{eqnarray}
Using the definition of the Fourier components $c_k=\sum_{n=1}^N y_n e^{-i 2
  \pi k x_n}$ this formula can be written as
\begin{equation}
\frac{\partial \chi^2}{\partial k_0} = -2 N\sum_{k=0}^{N-1} 
( B'_k c_k c_k^* + B'_k B_k c_k c_k^* ) + 32 \pi^4 \lambda \sum_{k=0}^{N-1} B_k
B'_k k^4 c_k c_k^* , \nonumber
\end{equation}
where we use $1/N \sum_{n=1}^N e^{i2\pi (k-l) x_n}=\delta_{kl}$, $1/N
\sum_{k=0}^{N-1} e^{i 2 \pi k(x_n-x_m)}=\delta_{nm}$ and $\delta_{nm}$ the
Kronecker delta. The factor $N$ can be avoided, if one scales $k_0\mapsto 2
\pi/N k_0$.

We write the minimum condition as $0= -F + \lambda G$. Then a simple relation
$\lambda= F/G$ results, relating lambda to $k_0$. The inversion of this
formula yields $k_0(\lambda)$.

\end{document}